\documentclass[%
 reprint,
superscriptaddress,
reprint,
 amsmath,amssymb,
 aps,
prb,
]{revtex4-2}

\usepackage{graphicx}% Include figure files
\usepackage{dcolumn}% Align table columns on decimal point
\usepackage{bm}% bold math
\usepackage{hyperref}
\hypersetup{
    colorlinks = true,
    citecolor = {blue},
    urlcolor = {blue}
}
\usepackage{color}
\definecolor{red}{rgb}{1,0,0}

\begin{document}

\preprint{APS/123-QED}

\title{The pairing symmetry in quasi-one-dimensional superconductor Rb$_2$Mo$_3$As$_3$}% Force line breaks with \\

\author{\v Ziga Gosar}
 \affiliation{Institute Jo\v zef Stefan, Jamova 39, SI-1000, Ljubljana, Slovenia}%Lines break automatically or can be forced with \\
 \affiliation{Faculty of mathematics and physics, University of Ljubljana, Jadranska 19, SI-1000, Ljubljana, Slovenia}%Lines break automatically or can be forced with \\

\author{Tina Arh}%
\affiliation{Institute Jo\v zef Stefan, Jamova 39, SI-1000, Ljubljana, Slovenia}%Lines break automatically or can be forced with \\
\affiliation{Faculty of mathematics and physics, University of Ljubljana, Jadranska 19, SI-1000, Ljubljana, Slovenia}%Lines break automatically or can be forced with \\

\author{Kevin Jakseti\v c}%
\affiliation{Institute Jo\v zef Stefan, Jamova 39, SI-1000, Ljubljana, Slovenia}%Lines break automatically or can be forced with \\

\author{Andrej Zorko}%
\email{andrej.zorko@ijs.si}
\affiliation{Institute Jo\v zef Stefan, Jamova 39, SI-1000, Ljubljana, Slovenia}%Lines break automatically or can be forced with \\
\affiliation{Faculty of mathematics and physics, University of Ljubljana, Jadranska 19, SI-1000, Ljubljana, Slovenia}%Lines break automatically or can be forced with \\

\author{Wenhao Liu}%
\affiliation{University of Texas at Dallas, 800 West Campbell Road Richardson, Texas}%

\author{Hanlin Wu}%
\affiliation{University of Texas at Dallas, 800 West Campbell Road Richardson, Texas}%

\author{Chennan Wang}
\affiliation{Laboratory for Muon Spin Spectroscopy, Paul Scherrer Institute, CH-5232 Villigen, Switzerland}

\author{Hubertus Luetkens}
\affiliation{Laboratory for Muon Spin Spectroscopy, Paul Scherrer Institute, CH-5232 Villigen, Switzerland}

\author{Bing Lv}
\affiliation{University of Texas at Dallas, 800 West Campbell Road Richardson, Texas}%

\author{Denis Ar\v con}%
\email{denis.arcon@ijs.si}
\affiliation{Institute Jo\v zef Stefan, Jamova 39, SI-1000, Ljubljana, Slovenia}%Lines break automatically or can be forced with \\
\affiliation{Faculty of mathematics and physics, University of Ljubljana, Jadranska 19, SI-1000, Ljubljana, Slovenia}


\date{\today}% It is always \today, today,

\begin{abstract}
Quasi-one-dimensional electron systems display intrinsic instability towards long-range ordered phases at sufficiently low temperatures. The superconducting orders are of particular interest as they can possess either singlet or triplet pairing symmetry and frequently compete with magnetism. Here we report on muon spin rotation and relaxation ($\mu$SR) study of Rb$_2$Mo$_3$As$_3$ characterised by one of the highest critical temperatures $T_{\rm c}=10.4$~K among quasi-one-dimensional superconductors. %The magnetic penetration depth $\lambda=669$~nm is compared to the coherence length $\zeta=3.4$~nm to obtain  the Ginzburg–Landau parameter $\kappa \approx 200$ and classify Rb$_2$Mo$_3$As$_3$ as a strong type-II superconductor in the clean limit. 
The transverse-field $\mu$SR signal shows enhanced damping below $T_{\rm c}$ due to the formation of vortex lattice. Comparison of vortex lattice broadening against single gap $s-$, $p-$ and $d-$wave models shows the best agreement for the  $s-$wave scenario but with the anomalously small superconducting gap, $\Delta_0$, to $T_{\rm c}$ ratio of $2\Delta_0/k_{\rm B}T_{\rm c}=2.74(1)$. The alternative nodal $p-$wave or $d-$wave scenarios with marginally worse goodness of fit would yield more realistic  $2\Delta_0/k_{\rm B}T_{\rm c}=3.50(2)$ and $2\Delta_0/k_{\rm B}T_{\rm c}=4.08(1)$, respectively, and thus they cannot be ruled out when accounting for the superconducting state in  Rb$_2$Mo$_3$As$_3$.  
\end{abstract}

\maketitle

%\tableofcontents

\section{Introduction}

Electron interactions in a one-dimensional (1D)  system turn any excitation into a collective one with characteristic  power-law dependencies of correlation functions, which is remarkably well described by the Tomonaga-Luttinger liquid (TLL) theory \cite{giamarchi1988, giamarchi2003, schulz2000fermi}.
The TLL theory also predicts that electron correlations in 1D trigger fluctuations in the spin-density-wave,  charge-density-wave or superconducting order parameters. In realistic quasi-1D systems the weak coupling between chains then stabilizes three-dimensional long-range order at finite temperature. 
There are only a handful examples of quasi-1D strongly correlated electron systems with competing insulating magnetically ordered  and unconventional superconducting states. For example, Bechgaard salts \cite{Jerome, brown} showing strong suppression of superconducting critical temperature, $T_{\rm c}$, by disorder \cite{Choi-PRB-2982}, the persistence of superconductivity  at fields that by far exceed the paramagnetic limit \cite{Lee-PRL-1997} and the absence of characteristic reduction in the NMR Knight shift  below $T_{\rm c}$ \cite{Lee-PRL} have been considered as candidates for the elusive triplet superconductivity. 
%An excellent model systems of such quasi-1D physics are Bechgaard salts \cite{Jerome, brown}, e.g. (TMTSF)$_2$PF$_6$ where TMTSF is tetramethyltetraselenafulvalene. 

%They are archetypal examples of strongly correlated electron systems with competing insulating magnetically ordered  and superconducting states. Remarkably, strong suppression of superconducting critical temperature, $T_{\rm c}$, by disorder \cite{Choi-PRB-2982}, the fact that superconductivity persists at fields that by far exceed the paramagnetic limit \cite{Lee-PRL-1997} and the absence of characteristic reduction in the NMR Knight shift \cite{Lee-PRL} below $T_{\rm c}$ suggest that (TMTSF)$_2$PF$_6$ may be a strong contender for the elusive triplet superconductivity. 

Recently,  A$_2$Cr$_3$As$_3$ \cite{Bao_PRX2015, Wu-PRB-2015, Pang-PRB2015, Tang-RbCr, Watson-PRL-2017, Zhi_PRB, Zhi-PRL2015, Pang-PRB-2015, Adroja-PRB-2015, Adroja_JPSJ-2017, Taddei-PRB-2017, xu2020coexistence, Yang-SciAdv-2021} and A$_2$Mo$_3$As$_3$ (A= K, Rb, Cs) \cite{MU-2018, Yang2019, Gosar-2020, Singh-PRB-2021, Taddei-arxiv-2022} have emerged as new candidates for such quasi-1D superconductors. Their structure comprises assembled Cr$_3$As$_3$ or Mo$_3$As$_3$  chains separated by alkali metals \cite{Bao_PRX2015, MU-2018}.  First principle calculations emphasise their quasi-1D electronic structure: the Fermi surface consists of two one-dimensional and one three-dimensional components \cite{Wu-PRB-2015, xu2020coexistence, Singh-PRB-2021, Yang2019}. The presence of 1D features is reflected in their highly anisotropic transport properties and TLL physics probed for example by nuclear magnetic resonance (NMR) \cite{Zhi_PRB, Gosar-2020, Yang-SciAdv-2021}. The emerging superconducting state shows many properties of unconventional superconductors, including the large specific heat jump at $T_{\rm c}$ and large upper critical fields exceeding the Pauli limit \cite{Bao_PRX2015, Tang-RbCr}, the absence of the Hebel-Slichter coherence peak and the power-law dependence of nuclear spin-lattice relaxation rate $1/T_1$ \cite{Zhi_PRB, Zhi-PRL2015, Gosar-2020}.  
Moreover, the transverse field muon spin relaxation ($\mu$SR) results on Cs$_2$Cr$_3$As$_3$  are more consistent with a nodal gap structure than an isotropic $s$-wave model for the superconducting gap, while the zero-field $\mu$SR relaxation is enhanced below $T_{\rm c}$ thus hinting to the triplet-type superconductivity \cite{Adroja-PRB-2015, Adroja_JPSJ-2017}.   
Therefore, the possibility of triplet superconductivity has been discussed in the literature for these two families of materials but the consensus about the pairing symmetry has not  been reached yet. 

A recent theoretical study suggests that A$_2$Cr$_3$As$_3$ with $T_{\rm c}\sim 6$~K is indeed an unconventional superconductor possibly hosting triplet superconductivity, but the related Mo-analogues A$_2$Mo$_3$As$_3$, with higher $T_{\rm c}\sim 10$~K, should be conventional multigap superconductors \cite{Singh-PRB-2021}.
Inelastic neutron scattering seems to corroborate this picture -- while antiferromagnetic spin fluctuations are present in both families, they are in the superconducting state of K$_2$Cr$_3$As$_3$ gapless, but gapped in K$_2$Mo$_3$As$_3$ \cite{Taddei-arxiv-2022}. In the latter case the gap opens below $\sim 6$~K for energies below $\sim 5$~meV. Although the data does not allow to unambiguously discriminate between nodal or nodeless gap functions, it has been argued that K$_2$Mo$_3$As$_3$ may  belong to conventional superconductors as theoretically suggested.   
That would imply the leading role of a three-dimensional Fermi surface for the occurrence of superconductivity. 
On the other hand,  power-law dependence of $^{87}$Rb $1/T_1$ in Rb$_2$Mo$_3$As$_3$  underlines the importance of 1D Fermi surface components carrying TLL as a parent state for the low-temperature unconventional superconductivity \cite{Gosar-2020}. 

To throw some additional light on  apparently contradicting experiments and theory we here report a $\mu$SR study of Rb$_2$Mo$_3$As$_3$. We estimate  the Ginzburg–Landau parameter $\kappa \approx 200$ that classifies Rb$_2$Mo$_3$As$_3$ as a strong type-II superconductor in the clean limit. More importantly, transverse field (TF) $\mu$SR allows us to determine the temperature dependence of the magnetic penetration depth. The  superconducting contribution to the TF $\mu$SR  relaxation
does not show any signs of saturation down to $T/T_{\rm c}\sim 0.14$, which is generally incompatible with a single-gap $s-$wave model. Furthermore, forcing this model of conventional superconductivity yields an anomalously small superconducting gap, $\Delta_0$, to $T_{\rm c}$ ratio $2\Delta_0/k_{\rm B}T_{\rm c}=2.74(1)$. On the other hand, almost equally good fits are obtained for the nodal  $p-$wave or $d-$wave scenarios with much  more realistic  $2\Delta_0/k_{\rm B}T_{\rm c}=3.50(2)$ and $2\Delta_0/k_{\rm B}T_{\rm c}=4.08(1)$, respectively. Present results do not definitely rule out the conventional $s-$wave scenario, but also keep the  door wide open for the unconventional nodal-type superconductivity in Rb$_2$Mo$_3$As$_3$.

\section{Experimental methods}

Preparation of polycrystalline Rb$_2$Mo$_3$As$_3$ followed the same steps as reported previously \cite{Gosar-2020}. Laboratory powder X-ray diffraction  suggested a phase pure sample with a hexagonal crystal lattice and  space group  $P\overline{6}m2$. High sample quality was further confirmed by dc magnetic susceptibility measurements in the zero-field cooling protocol at $\mu_0H = 1$~mT, which showed  bulk superconductivity below $T_{\rm c}= 10.4$~K  with a diamagnetic shielding fraction of almost 100\%. To avoid any possible sample degradation we strictly avoided the exposure of  samples to air at all stages - sample handling and packing was done in Ar-filled glove box with a control atmosphere where O$_2$ and H$_2$O contents were below 0.1~ppm and sample was transported between experimental sites in a  glass tube sealed under high vacuum.  

All $\mu$SR measurements were conducted on the  GPS instrument at the Paul Scherrer Institut (PSI), Switzerland  \cite{Amato-2017}.   
Polycrystalline sample in the form of a pellet with dimensions 5~mm in diameter and 4~mm thick was sealed between  two layers of kapton tape in a glovebox and then glued to a fork sample holder made of copper. $\mu$SR measurements were done in veto mode, thus minimizing the background signal to below $\sim 5$\}. Measurements of $\mu$SR signal were performed  in the zero-filed (ZF), longitudinal field (LF) and TF geometries \cite{yaouanc2011muon}. Stray fields from the Earth or from the neighbouring instruments were cancelled in ZF measurements with adaptive compensation coils. The majority of the TF $\mu$SR experiments were undertaken in a magnetic field of 100~mT, which is above $H_{\rm c1}$ and well below $H_{\rm c2}$. 
 In our measurements we  thermalised the sample at each temperature  prior the start of the measurement and used high statistics with about 20 million events for each dataset.
Raw $\mu$SR data were analysed using \texttt{musrfit} software \cite{musrfit}. 

\section{Results}

\begin{figure}[b]
\includegraphics[width=1.0\linewidth]{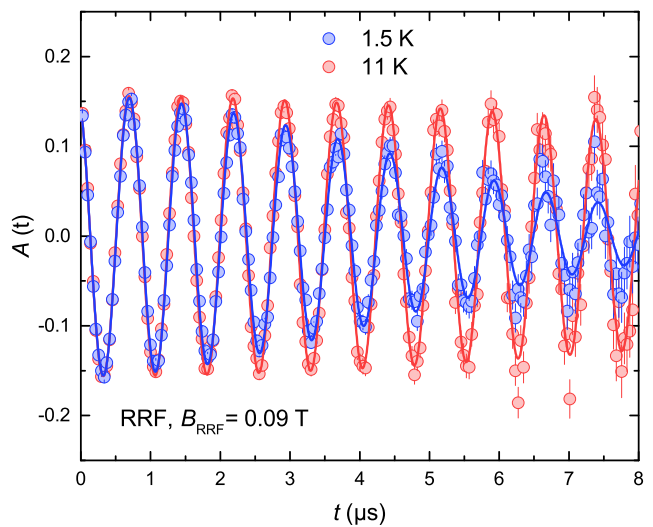}% Here is how to import EPS art
\caption{\label{fig:TFsignal} Transverse field (TF) $\mu$SR measurements of Rb$_2$Mo$_3$As$_3$ in a field of 100~mT at temperatures $T=11$~K (red), i.e., above $T_{\rm c}=10.4$~K and $T=1.5$~K (blue). Solid red and blue lines are fits to Eq.~\ref{eq:TFmSR}. The plot is shown in rotating reference frame (RRF) with the frequency that corresponds to 90~mT. }
\end{figure}

To determine the temperature dependence of the superfluid density, we first measured TF $\mu$SR above and below $T_{\rm c}$ using the standard field cooling (FC) thermal protocol. Characteristic TF $\mu$SR asymmetries taken at $T=11$~K (just above $T_{\rm c}$) and at $T=1.5$~K (well below $T_{\rm c}$) are compared in Fig.~\ref{fig:TFsignal}. The relaxation of TF $\mu$SR signal  above $T_{\rm c}$ is very weak and  mainly originates from the dipolar coupling to small static nuclear moments. Below $T_{\rm c}$, the relaxation of TF $\mu$SR signal is significantly enhanced. This is a signature of the vortex lattice formation leading to inhomogeneous field distribution at the implanted muon sites. 

Fourier transform of the TF $\mu$SR data yields spectra with a sharp line at the applied field and a broader component shifted to lower fields (not shown). The former, much less intense signal is likely due to muons stopping  either in a parasitic non-superconducting phase or in the sample holder, 
while the broader  component shifted to lower fields  originates from the sample in the superconducting state. Both components can be well described by a Gaussian lineshape. The expected asymmetric lineshape broadening of the superconducting component \cite{Maisuradze_2009} was also tested by assuming skewed Gaussian lineshape \cite{musrfit} but the resulting  asymmetry in the  widths was very small.   We thus analyze  the TF $\mu$SR data at all temperatures with two components, both experiencing  Gaussian relaxation,
\begin{eqnarray}
    A(t) = && A[ (1-f)\exp\left[-\frac{1}{2}(\sigma t)^2\right]\cos (2\pi\nu t +\varphi) \nonumber\\
    && +F \exp\left[-\frac{1}{2}(\sigma_b t)^2\right]\cos (2\pi\nu_b t +\varphi) ]\,  .
    \label{eq:TFmSR}
\end{eqnarray}
Here $\sigma$ and $\sigma_b$ are the Gaussian relaxation rates for the sample and background components, respectively,  $\nu$ and $\nu_b$ are the respective muon precession frequencies, $\varphi$ is the phase given by the detector geometries and $F=0.148$ is the fraction of the background signal. The latter was determined for the data collected at $T=1.5$~K and then kept constant for all other temperatures. The  relaxation rate of the background component $\sigma_{\rm b}=0.158$~$\mu$s$^{-1}$ and its resonance field $2\pi\nu_b/\gamma_\mu=99.887$~mT were also fixed. 

\begin{figure}[t]
\includegraphics[width=1.0\linewidth]{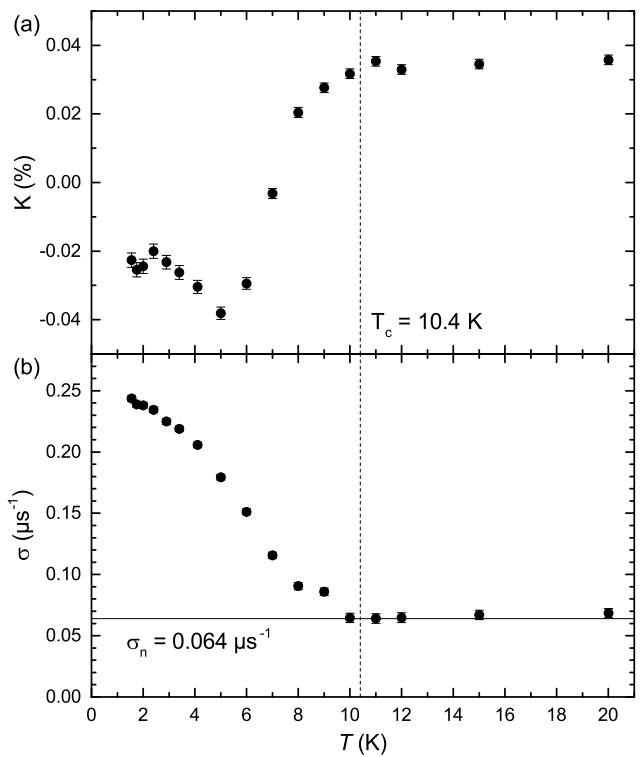}% Here is how to import EPS art
\caption{\label{fig:TFKnight} (a) The temperature dependence of  the experimental frequency shift, $K$, in the quasi-one-dimensional Rb$_2$Mo$_3$As$_3$ superconductor obtained from  the fits of TF $\mu$SR data with Eq.~\ref{eq:TFmSR}.  (b) The temperature dependence of TF $\mu$SR Gaussian relaxation rates, $\sigma$. The magnetic field was set to 100~mT and the data corresponds to  the measurements collected after the field-cooled protocol.  The vertical dashed line marks $T_{\rm c}=10.4$~K, while the solid horizontal line in (b) indicates the nuclear contribution to the TF $\mu$SR relaxation, $\sigma_{\rm n}=0.064$~$\mu$s$^{-1}$.  }
\end{figure}

These assumptions allowed us to determine the temperature dependencies of the experimental frequency shift, $K=(\nu-\nu_{\rm b})/\nu_{\rm b}$, measured against $\nu_{\rm b}$ and the corresponding relaxation rate, $\sigma$, of the Rb$_2$Mo$_3$As$_3$  sample in the FC experiment (Fig.~\ref{fig:TFKnight}). 
The $\mu$SR spectrum of the superconducting component starts to diamagnetically shift at $T_{\rm C}=10.4$~K (Fig.~\ref{fig:TFKnight}a). 
%However, the precise temperature dependence of $K(T)$ below $T_{\rm C}$ depends on the cooling protocol and can be as large as $K=-0.8$\% in the zero-field cooling (ZFC) protocol (not shown). In our main study we followed the FC protocol. In these experiments, $K(T)$ initially shows a diamagnetic shift below $T_{\rm C}$ (Fig. \ref{fig:TFKnight}a). However, $K(T)$ seems to   slightly reverse its trend below $T\sim 6$~K. 
The muon relaxation rate gets  enhanced just below $T_{\rm c}$ and $\sigma (T)$ monotonically increases  with lowering temperature (Fig.~\ref{fig:TFKnight}b), but  does never really saturate even at $T/T_{\rm c}\approx 0.14$. 
%At $T\sim 6$~K there seems to be an inflection point in the relaxation rate which does never really saturates even at $T/T_{\rm c}\approx 0.14$. Curiously, for the TF $\mu$SR experiments taken at $T=1.5$~K, $\sigma$ monotonically increases with magnetic field (inset to Fig. \ref{fig:TFKnight}b). 

The relaxation rate $\sigma$ of muons implanted in sample has two contributions: the  temperature-dependent contribution from the vortex lattice, which is dominant in the superconducting phase, $\sigma_{\rm sc}$, and the smaller temperature-independent contribution from the nuclear dipole moments, $\sigma_{\rm n}$. The total Gaussian relaxation rate is then given by $\sigma =\sqrt{\sigma_{\rm sc}^2+\sigma_{\rm n}^2}$ \cite{Adroja_JPSJ-2017, Adroja-PRB-2015, adroja2018multigap, Biswas-PRB2018, prozorov2006}. Above $T_{\rm c}$, $\sigma_{\rm sc}=0$, which allows us to determine $\sigma_{\rm n}=0.064$~$\mu$s$^{-1}$.

As the applied TF field of $100$~mT is much smaller than the upper critical field $\mu_0H_{c2}\approx 28$~T, we can use the calculated value of $\sigma_{\rm sc}=0.24$~$\mu$s$^{-1}$ at $T=1.5$~K to estimate the effective penetration depth, $\lambda=\sqrt{0.0609\gamma_\mu\Phi_0/\sigma_{\rm sc}}=669$~nm \cite{Brandt-PRB}. Here $\Phi_0$ is the magnetic flux quantum and $\gamma_\mu$ is the muon gyromagnetic ratio. Taking into consideration also the  coherence length $\zeta=3.4$~nm, estimated from $H_{\rm c2}$, this yields the   Ginzburg–Landau parameter $\kappa=\lambda/\zeta  \approx 200$ and classifies Rb$_2$Mo$_3$As$_3$ as a strong type-II superconductor.
We note that the above expression for the effective penetration depth holds for $0.13/\kappa^2\ll (H/H_{c2}) \ll 1$ and $\kappa \gg 70$ \cite{Brandt-PRB}, which is well justified in our  TF experiments. 
Finally, the carrier mean free path calculated based on  resistivity data \cite{Gosar-2020} is $l_{\rm eff}\approx 20$~nm, thus the studied compound seems to be  in the clean limit. 

Rb$_2$Mo$_3$As$_3$ is a quasi-1D superconductor and thus the  anisotropy of the penetration depths $\gamma_\lambda = \lambda_c/\lambda_{ab}$ could be present. Here $\lambda_c$ and $\lambda_{ab}$ are the penetration depths along the chain and in the plane perpendicular to the chains, respectively. 
 In such cases, the  effective penetration depth may differ from  $\lambda_{ab}$ and $\lambda_c$ by a factor close to 1 \cite{adroja2018multigap}, but this should not affect the overall temperature dependence of $\sigma_{\rm sc}$.
In any case, the  anisotropy of the iron-pnictide superconductors is usually found to be small \cite{Yuan-Nature2009} and thus we speculate that the effective $\lambda$ is not much different from $\lambda_c$ and $\lambda_{ab}$. Therefore, we proceed with our analysis using the effective $\lambda(T)$ which is inversely proportional to the square root of $\sigma_{\rm sc}$ and  relate it to the superfluid density and its temperature dependence \cite{prozorov2006},
\begin{equation}
    \frac{\sigma_{\rm sc}(T)}{\sigma_{\rm sc}(0)} = 1+\frac{1}{\pi} \int_0^{2\pi}\int_{\Delta(T,\phi)}^{\infty} \frac{\partial f}{\partial E}\frac{EdEd\phi}{\sqrt{E^2-\Delta^2(T,\phi)}}\, .  
     \label{eq:sigma}
\end{equation}
Here $f=[1+\exp (-E/k_{\rm B}T)]^{-1}$ is the Fermi function. The superconducting gap  is temperature and angular dependent through $\Delta(T,\phi)=\Delta_0\delta(T/T_{\rm c})g(\phi)$ \cite{Pang-PRB2015}. $\Delta_0$ is the  maximum gap value at $T=0$. Following the literature, we use the BCS expression for the temperature dependence of the superconducting gap $\delta(T/T_{\rm c})=\tanh [1.82\cdot(1.018(T_{\rm c}/T-1))^{0.51})]$ \cite{Carrington-2003}. With $\phi$ as an azimuthal angle along the Fermi surface, $g(\phi)$ takes into account the angular dependence of the superconducting gap. In our modeling we assumed isotropic ($s-$wave) $g(\phi)=1$, $g(\phi)=\vert \cos (\phi)\vert$ for the $p-$wave, and  $g(\phi)=\vert \cos (2\phi)\vert$ for the $d-$wave nodal gaps. 

\begin{figure}[htbp]
\includegraphics[width=1.0\linewidth]{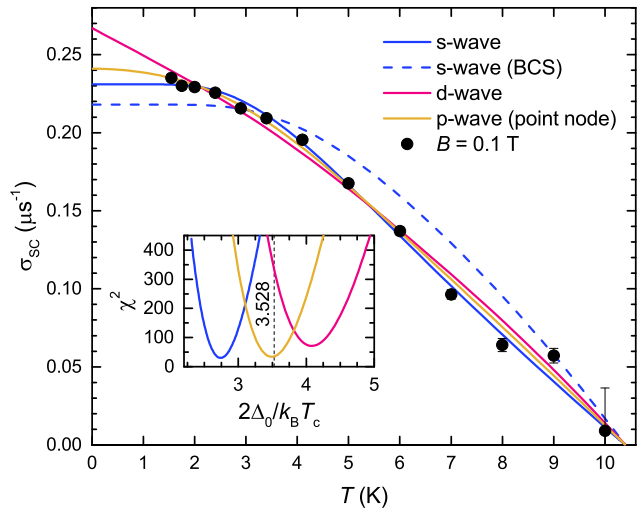}% Here is how to import EPS art
\caption{\label{fig:sigma} The temperature dependence of  the  TF $\mu$SR  superconducting relaxation rate, $\sigma_{\rm sc}$ (black circles). Dashed blue line is the calculated $\sigma_{\rm sc}(T)$ for the $s-$wave scenario with the BCS value of the superconducting gap, $\Delta_0$, to $T_{\rm c}$ ratio $2\Delta_0/k_{\rm B}T_{\rm c}=3.528$. Solid blue, orange and pink lines correspond to $s-$, and nodal $p-$ and $d-$waves with the optimized $2\Delta_0/k_{\rm B}T_{\rm c}$ values, respectively. Inset: Variation of the goodness of fit $\chi^2$ against $2\Delta_0/k_{\rm B}T_{\rm c}$ values for all three single-gap models.}
\end{figure}

The temperature dependence of $\sigma_{\rm sc}(T)$ is in Fig.~\ref{fig:sigma} compared against different single-gap models computed from Eq.~\ref{eq:sigma}. It is immediately evident that the conventional single-gap $s-$wave model with the weak-coupling BCS value of the superconducting gap to $T_{\rm c}$ ratio $2\Delta_0/k_{\rm B}T_{\rm c}=3.528$ completely fails in describing the data. The goodness of fit, $\chi^2$, is significantly improved (from previous $\chi^2>400$  to $\chi^2=30.2$) if the superconducting gap is reduced to $\Delta_0/k_{\rm B} = 14.2(1)$~K so that $2\Delta_0/k_{\rm B}T_{\rm c}=2.74(1)$ (inset to Fig.~\ref{fig:sigma}). We stress that this gap value is significantly smaller than it is predicted by the conventional BCS value. We also tested anisotropic $s-$wave (but nodeless) model, but this model also converges to values of $2\Delta_0/k_{\rm B}T_{\rm c}$ much smaller than the BCS value.  Remarkably, the point nodal $p-$wave model gives only marginally worse $\chi^2=33.8$, but for $\Delta_0/k_{\rm B}=18.2(1)$~K and thus more reasonable $2\Delta_0/k_{\rm B}T_{\rm c}=3.50(2)$. This model also agrees better with the observation that $\sigma_{\rm sc}(T)$ does not show any signs of saturation down to $T/T_{\rm c}\approx 0.14$. Finally, we tested also the line nodal $d-$wave model that shows somewhat worse $\chi^2=71.1$ for the optimised $2\Delta_0/k_{\rm B}T_{\rm c}=4.08(1)$. 

As the Fermi surface of Rb$_2$Mo$_3$As$_3$ comprises two quasi-1D and one three-dimensional components, we tried also the two-gap $s+s$-wave model. We assumed a weighted sum of two normalised superfluid densities calculated from Eq.~\ref{eq:sigma} to obtain
\begin{equation}
    \sigma_{\rm sc}(T)=w\sigma_{\rm sc}(\Delta_1,T)+
    (1-w)\sigma_{\rm sc}(\Delta_2,T)\, .
    \label{eq:sigma2}
\end{equation}
To minimise the number of free parameters,  the first component with the BCS superconducting gap $2\Delta_1/T_{\rm c}=3.528$  was fixed. Its weight, $w$, and the value of $2\Delta_2/T_{\rm c}$ for the second component with a smaller superconducting gap $\Delta_2$  was used to minimise $\chi^2$. The parameter optimisation spontaneously converged to a very small weight of the BCS-component, i.e., to $w=0.1$ and $2\Delta_2/k_{\rm B}T_{\rm c}=2.67$ ($\chi^2=29.6$). The marginal weight for the component with the larger $\Delta_1$ shows that the two-gap model is essentially very similar to the solution with a single anomalously small $s-$wave gap as discussed above. %It thus appears that  the two-gap $s+s$-wave model may not adequately explain TF $\mu$SR data in Rb$_2$Mo$_3$As$_3$. 

\begin{figure}[t]
\includegraphics[width=1.027\linewidth]{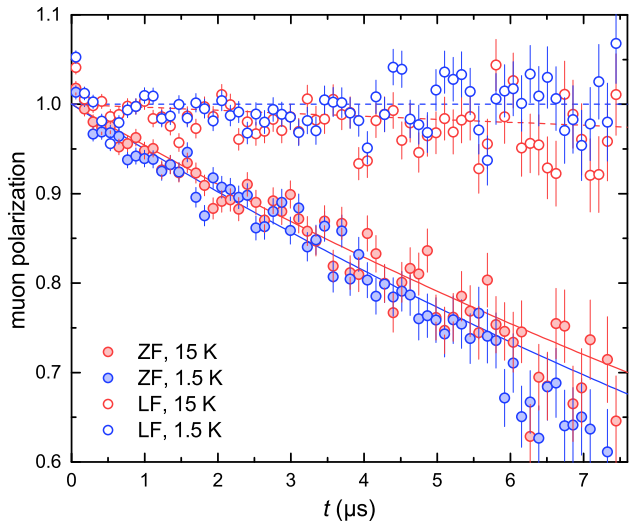}% Here is how to import EPS art
\caption{\label{fig:ZFmuSR} The time dependence of  the zero-field (ZF) $\mu$SR (solid circles) and weak 10~mT longitudinal-field (LF) (open circles) muon polarization collected  above $T_{\rm c}$ at $T=15$~K (red symbols) and well below $T_{\rm c}$ at $T=1.5$~K (blue symbols). Solid lines are fits of the  ZF  and dashed lines of the LF $\mu$SR data to a simple exponential decay function (Eq.~\ref{eq:ZFmuSR}).}
\end{figure}

Since the $p-$wave single gap model appears as a strong candidate to fit TF $\mu$SR data, we next focus on ZF $\mu$SR measurements to test possible time-reversal symmetry-breaking superconducting state. In Fig.~\ref{fig:ZFmuSR} we compare ZF-$\mu$SR data collected above $T_{\rm c}$ at $T=15$~K to data accumulated well below $T_{\rm c}$ at $T=1.5$~K. For both temperatures the data can be reasonably fit to a simple exponential decay function
\begin{equation}
    A(t)=A_0\exp(-\lambda_{\rm ZF}t)\, .
    \label{eq:ZFmuSR}
\end{equation}
Here $A_0$ is the initial asymmetry, while $\lambda_{\rm ZF}$ is the muon spin relaxation rate due to local magnetic fields. Fitting  ZF $\mu$SR  data to  the expression Eq.~\ref{eq:ZFmuSR} shows that $\lambda_{\rm ZF}$ is slightly enhanced in the superconducting state:  $\lambda_{\rm ZF}=0.047(1)$~$\mu$s$^{-1}$ at $15$~K increases to $\lambda_{\rm ZF}=0.052(1)$~$\mu$s$^{-1}$ at $1.5$~K. The muon spin relaxation is quickly suppressed already in small longitudinal fields of 10~mT (Fig.~\ref{fig:ZFmuSR}) implying slow local field fluctuations or even quasi-static local fields, as usually originating from nuclei.  
Even if the unknown background contribution is added to Eq. \ref{eq:ZFmuSR}, the extracted value of $\lambda_{\rm ZF}$ for $T=1.5$~K would still be larger compared to that for the  $T=15$~K dataset. For example, with 5\% of background included,  as it might be expected for our experimental setup, 
$\lambda_{\rm ZF}$  changes to 0.050(2) $\mu$s$^{-1}$ and 0.055(2) $\mu$s$^{-1}$ for $T=1.5$~K and $T=15$~K, respectively.
The minute enhancement in  $\lambda_{\rm ZF}$ in the superconducting state seems to be within the  uncertainty of current experiments and does not allow to unambiguously confirm the conjecture of time reversal symmetry-breaking superconducting state.

\section{Discussion and conclusions}

The main experimental findings of the present $\mu$SR study of the quasi-1D superconductor Rb$_2$Mo$_3$As$_3$ may be summarised as follows: (i) the vortex-lattice contribution to the Gaussian relaxation rate does not show any saturation even for $T/T_{\rm c}\approx 0.14$, (ii) the TF $\mu$SR frequency shift shows the diamagnetic shift below $T/T_{\rm c}$, but also an anomaly at  $\sim 6$~K, (iii) the ZF relaxation seems to be marginally enhanced at $T=1.5$~K compared to that measured above $T_{\rm c}$ at  $T=15$~K. Should the conventional superconductivity scenario  apply to Rb$_2$Mo$_3$As$_3$, then the small relative gap value, $2\Delta_0/k_{\rm B}T_{\rm c}=2.74(1)$, needs to be explained. It also appears that the modeling of TF $\mu$SR relaxation rate using nodeless anistropic $s-$wave or two-gap $s+s-$wave models does not resolve the issue of anomalously small $\Delta_0$. We note that the similarly small ratio   $2\Delta_0/k_{\rm B}T_{\rm c}$ as found by $\mu$SR has been previously deduced also from our $^{87}$Rb NMR data \cite{Gosar-2020}.  As possible explanation for the anomalous $2\Delta_0/k_{\rm B}T_{\rm c}< 3.528$ extracted from NMR data, we previously considered the gap averaging due to impurity scattering and a weighted average of the relaxation in the vortex core and in the inter-vortex region. 

However, the present $\mu$SR analysis shows that the nodal-superconductivity scenario is also possible for Rb$_2$Mo$_3$As$_3$. The temperature dependence of $\sigma_{\rm sc}$ and the minute enhancement of $\lambda_{\rm ZF}$ in the superconducting state are both compliant with the $p-$wave superconductivity. The enhancement  in $\lambda_{\rm ZF}$ below $T_{\rm c}$ by $5\cdot 10^{-3}$~$\mu$s$^{-1}$ is  similar to that observed in the sister  compound Cs$_2$Cr$_3$As$_3$ and 20 times larger than in  K$_2$Cr$_3$As$_3$ \cite{Adroja_JPSJ-2017}. Despite the similarity in the ZF $\mu$SR data between the Mo- and Cr-based families, it has to be stressed though that more detailed measurements are needed to unambiguously confirm enhancement in $\lambda_{\rm ZF}$ in the case of Rb$_2$Mo$_3$As$_3$. If the nodal-superconductivity scenario for Rb$_2$Mo$_3$As$_3$ is correct, then the anomaly in the TF $\mu$SR frequency shift below $\sim 6$~K may indicate a weak internal field due to the presence of the triplet superconducting component. Interestingly, we note that inelastic neutron scattering data indicate the opening of the superconducting gap below $\sim 6$~K in K$_2$Mo$_3$As$_3$, although it has a similar $T_{\rm c}\approx 10$~K \cite{Taddei-arxiv-2022} as Rb$_2$Mo$_3$As$_3$ studied here.   

While NMR, NQR and now also $\mu$SR techniques seem to be  consistent when probing the superconducting state of Rb$_2$Mo$_3$As$_3$, one cannot overlook the apparent inconsistencies  with the inelastic neutron scattering \cite{Taddei-arxiv-2022} or with first principle calculations \cite{Singh-PRB-2021}. For example, the superconducting gap deduced from all three  local-probe methods is around $\Delta_0\approx 1.2$~meV, which is much less than 3.1~meV extracted from the neutron scattering. As magnetic resonance methods are very sensitive to low-energy excitations, they  offer {\em a priori} better energy resolution in this energy range. More precise  $\mu$SR experiments extended to $T/T_{\rm c}\ll 0.1$ are therefore desired to ultimately address the symmetry of the superconducting order parameter in the Rb$_2$Mo$_3$As$_3$ quasi-1D superconductor.

\begin{acknowledgments}
DA wishes to acknowledge the support of the Slovenian research agency through research program No. P1--0125 and research projects N1--0220 and J1--3007. AZ acknowledges the support of the Slovenian research agency through the research project N1--0148.
This work at University of Texas at Dallas is supported by US Air Force Office of Scientific Research Grant No. FA9550--19--1--0037 and National Science Foundation (NSF)--DMREF--1921581.

\end{acknowledgments}

%\appendix

%\section{Appendixes}
%....

%\bibliography{Rb233-refs}% Produces the bibliography via BibTeX.

\providecommand{\noopsort}[1]{}\providecommand{\singleletter}[1]{#1}%

\end{document}